\documentclass[sigconf,nonacm]{acmart}

\usepackage{amsmath,amsfonts}
\usepackage{algorithmicx}
\usepackage{algorithm}
\usepackage{algpseudocode}
\usepackage{array}
\usepackage{makecell}
\usepackage{url}
\usepackage{verbatim}
\usepackage{graphicx}
\usepackage{enumitem}
\usepackage{xcolor}
\usepackage{tikz}
\usetikzlibrary{arrows.meta,positioning,fit,shapes.geometric,calc,shadows.blur,backgrounds}

\definecolor{figblue}{RGB}{70,118,170}
\definecolor{figgreen}{RGB}{76,140,102}
\definecolor{figorange}{RGB}{196,124,62}
\definecolor{figgray}{RGB}{88,96,105}
\definecolor{commentpurple}{RGB}{112,67,148}

\newcounter{papercomment}

\DeclareRobustCommand{\ourmethod}{\textsc{LSM-Vec}}

\AtBeginDocument{%
  }

\definecolor{aoenglish}{rgb}{0.8,0.2,0.0} 
\newcounter{dingheng}
\numberwithin{dingheng}{section}

\begin{document}

\title{{\ourmethod}: A Memory-Efficient Disk-Based System for Large-Scale Dynamic Vector Search}

\settopmatter{authorsperrow=4}

\author{Dingheng Mo}
\affiliation{%
  \institution{Nanyang Technological University}
  \country{Singapore}}
\email{dingheng001@e.ntu.edu.sg}

\author{Shurui Zhong}
\affiliation{%
  \institution{Nanyang Technological University}
  \country{Singapore}}
\email{SHURUI001@e.ntu.edu.sg}

\author{Yingzhuo Qian}
\affiliation{%
  \institution{Google}
  \country{United States}}
\email{yingzhuoq@google.com}

\author{Siqiang Luo}
\affiliation{%
  \institution{Nanyang Technological University}
  \country{Singapore}}
\email{siqiang.luo@ntu.edu.sg}

\begin{abstract}

RAG and AI retrieval have made vector approximate nearest-neighbor (ANN) search a key retrieval operation in modern data systems. These workloads are both large and dynamic. Vector collections grow until index state no longer fits in DRAM and must reside on disk, and every new record arrives as an update the index must absorb while queries continue to run. Existing dynamic designs on disk leave two architecture-level gaps. First, graph connections and vector data follow different access and update paths, yet many disk-based indexes store them as one physical record, so a graph-only update rewrites vector data it does not modify. Second, keeping an index fresh gives memory a role beyond caching, because these systems either hold recent updates in in-memory indexes or reserve memory to maintain a mutable disk layout.

We present {\ourmethod}, a disk-based dynamic vector search system built on a decoupled architecture. {\ourmethod} stores the HNSW base layer in a graph-oriented LSM-tree that absorbs fine-grained edge changes out of place, and keeps vector data in a separate paged store. To keep this design query-efficient within a small memory budget, we design in-memory tree-based sketches that route each query to an entry close to its target on disk, so the resident hierarchy stays compact. We further introduce HNSW-based reordering, which places vector data that queries evaluate together on nearby disk pages and reduces scattered reads. On streaming SIFT and SPACEV workloads with up to a hundred million vectors, {\ourmethod} requires up to 6.2\(\times\) less memory and 7.6\(\times\) less disk than FreshDiskANN and SPFresh, and writes up to 15\(\times\) fewer bytes per insertion. At the hundred-million scale, where SPFresh's index outgrows memory and FreshDiskANN exhausts it, {\ourmethod} keeps its tail latency stable throughout the update stream, 2.3--3.8\(\times\) below SPFresh's at comparable recall.

\end{abstract}




\maketitle

\section{Introduction}\label{sec:intro}

\vspace{1mm}
Given a collection of vectors and a query vector \(q\), vector approximate nearest-neighbor search, or simply approximate nearest-neighbor (ANN) search, aims to find the top-\(k\) closest vectors to the query \(q\), where the closeness is defined by a metric distance such as Euclidean distance. ANN search has become a key retrieval operation for modern AI applications. In retrieval-augmented generation (RAG), large language models use ANN search to fetch external knowledge before generation~\cite{lewis2020retrieval,guu2020retrieval,izacard2023atlas,borgeaud2022improving}. Recommendation and multimodal retrieval map unstructured records into embedding spaces where semantic similarity becomes a search operation~\cite{covington2016deep,radford2021learning}. As a result, vector search is no longer a standalone retrieval module. It is increasingly integrated into data systems that serve RAG and AI retrieval workloads. Industrial data platforms reflect this trend: Databricks, Snowflake, and MongoDB now expose vector search as a native data-system capability for RAG and AI retrieval over platform-managed data~\cite{databricks2026aisearch,snowflake2026cortexsearch,mongodb2026vectorsearch}.

This trend is reshaping real-world vector workloads in two directions at once, toward larger scale and higher update rates. Vector datasets keep growing, and every new record becomes an index update. Managed platforms are sized accordingly. On Google Cloud, managed vector search is sized for billions of embeddings~\cite{googlecloud2024vertexvectorsearch}, and newly inserted vectors are expected to become queryable within seconds~\cite{googlecloud2026streamingupdates}. At this scale, keeping the full graph and vector data in DRAM becomes unaffordable, so the index must reside on disk and continue to serve queries under continuous updates.

\begin{figure*}[!t]
    \vspace{-3mm}
    \centering
    \makebox[\textwidth][c]{\includegraphics[width=1.06\textwidth]{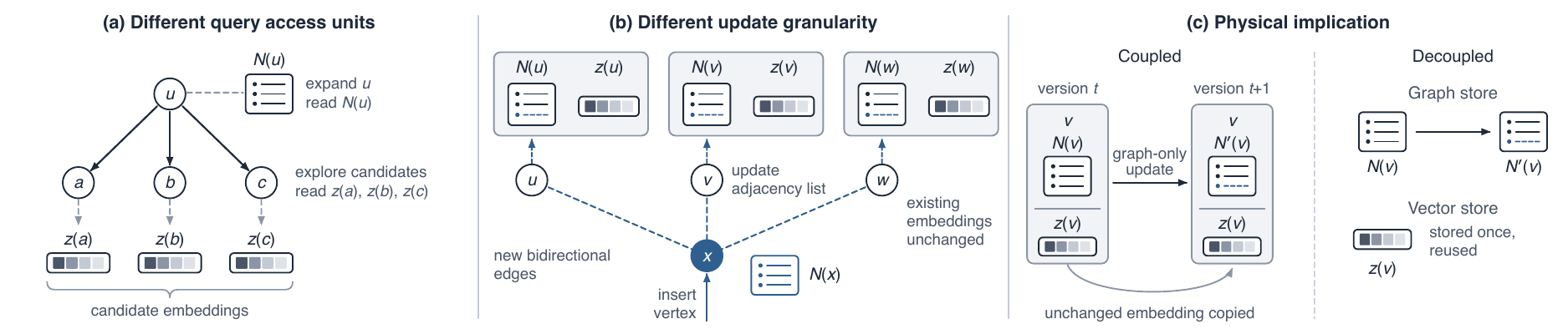}}
    \vspace{-3mm}
    \caption{Graph and vector data follow different query and update paths. (a) Expanding vertex \(u\) reads its adjacency list \(N(u)\), but evaluates the vector data of discovered neighbors. (b) Inserting \(x\) creates \(N(x)\) and \(z(x)\), while reciprocal linking and pruning may change existing adjacency lists without changing their vector data. (c) A coupled record may copy stable vector data during a graph-only update, whereas separated storage updates adjacency data independently.}
    \vspace{-3mm}
    \label{fig:access-update-units}
\end{figure*}

\vspace{1mm}
\noindent\textbf{Existing methods: Static graph layout and dynamic updates.}
Early work on searching vectors beyond DRAM capacity led to static graph indexes on disk. DiskANN~\cite{jayaram2019diskann} is a representative design: it builds a high-quality graph offline and serves queries from disk. As AI data-system workloads become more dynamic, recent work has turned to update-aware designs that keep vector indexes fresh as new data arrive.
FreshDiskANN~\cite{singh2021freshdiskann} adopts an out-of-place update architecture, buffering recent updates in temporary in-memory indexes and periodically merging them into a disk-based index. This design makes updates visible without requiring an immediate full rebuild, but queries must probe both the persistent index and the temporary update structures before merging completes. In contrast, SPFresh~\cite{xu2023spfresh} applies updates directly to a primary partitioned disk index. Although this eliminates the need for a separate update index, it requires the system to continuously maintain the partitioning structure. Despite these advances, existing designs still exhibit two fundamental architectural limitations.

\vspace{1mm}
\noindent\textbf{Gap 1: Graph and vector data follow different access paths, hence questioning the existing designs storing them in a bundle.}
In a graph-based ANN index, each vector is represented by a graph vertex that stores both its embedding and its outgoing edges. Many disk-based systems map this logical vertex directly to one physical record. DiskANN, for example, stores each full-precision vector together with its graph neighborhood on disk~\cite{jayaram2019diskann}. However, this can be inefficient because accesses to graph structures and vector embeddings are not synchronized during update and search.

The update path shows the first mismatch, illustrated in Figure~\ref{fig:access-update-units}(b). Inserting a new vector into HNSW creates its adjacency lists and also adds reciprocal edges in the neighborhoods it joins. By contrast, the existing vertices whose adjacency lists are modified keep their vector data unchanged. If graph connections and vector data are maintained in one disk block, a graph-only change can still read, copy, or rewrite vector bytes that the update does not modify.

The query path has its own mismatch, illustrated in Figure~\ref{fig:access-update-units}(a). When search expands a vertex \(u\), it reads \(u\)'s adjacency list and discovers neighboring vertices. At that point, the query needs the neighbors' vector data for vector-distance evaluation, not necessarily their adjacency lists. Graph access is therefore centered on the expanded vertex, while vector access is centered on the candidates it reveals. Thus, when each vertex's adjacency list is stored together with its own vector data, reading the neighbors' vector data also brings in their adjacency lists, which the expansion does not use.

This layout does help when a query reaches a vertex, because a single read then yields both of its components. Each operation above, however, uses one component at a time, so storing them together forces an insertion to move the vector data of the neighbors whose edges it changes, and an expansion to read the adjacency lists of the candidates it only needs to evaluate. A dynamic vector search system on disk should therefore keep graph connections and vector data in separate stores, as sketched in Figure~\ref{fig:access-update-units}(c). Each path then moves only the data it uses, and the vector store can be organized around the candidates an expansion evaluates, so that they are read from few pages. Section~\ref{sec:eval-ablation} isolates this choice: storing each vector with its adjacency list makes queries 3.4--3.6\(\times\) slower and insertions 3.0--3.3\(\times\) slower at equal recall.

\vspace{1mm}
\noindent\textbf{Gap 2: Dynamic updates give memory an architectural role, hence reducing memory usage is critical.}
Dynamic vector search on disk still gives memory a central role beyond caching. The key issue is architectural: existing dynamic methods must reserve memory for dynamic index state, either to expose recent updates to queries or to preserve a mutable disk layout.

FreshDiskANN~\cite{singh2021freshdiskann} maintains its dynamic state at the level of a complete ANN index. Its short-term in-memory indexes must support search until the next merge, so they cannot be reduced to append buffers. Merging more frequently would shrink them, but each merge is expensive. StreamingMerge folds the accumulated updates into a large graph index on disk, and folding into a proximity graph re-runs search and pruning rather than merging sorted runs, so its cost follows the index rather than the updates it absorbs. The merge interval must therefore remain large, and unmerged updates stay in memory throughout.

SPFresh~\cite{xu2023spfresh} reaches a different point in the same design space. By applying updates in place, it removes the in-memory update index altogether. However, this design still needs memory of its own, both for summaries that route queries to promising partitions and for metadata that tracks the partitioning structure as new vectors arrive. Thus, memory is used to maintain the organization of a mutable disk index rather than to hold a secondary update index.

The two designs place the dynamic state in different structures, but in both, memory holds state that grows with the index and its update volume rather than with the query working set. Moving vector data to disk therefore does not by itself bound memory, because the mechanism that keeps the index fresh needs memory of its own. A dynamic vector search system on disk should instead keep memory for compact routing state and caches, whose size is a configuration choice rather than a consequence of the update architecture.

\vspace{1mm}
\noindent\textbf{Our solution: A decoupled dynamic vector search system.}
To address these limitations, we propose {\ourmethod}, a disk-based dynamic vector search system built on a decoupled architecture. {\ourmethod} separates graph connections from vector data, so each path can use a storage layout matched to its access and update behavior. The graph path stores the HNSW base layer on disk in a graph-oriented LSM-tree~\cite{mo2025aster}, where fine-grained edge changes are absorbed out of place. The vector path stores fixed-length vector data in a separate paged store, where inserts append to pages and queries batch their reads by page.

This separation addresses the first limitation directly. An edge change is written to the graph store alone, so no vector byte is read or rewritten to record it, and the candidates discovered by an expansion are evaluated from their vector pages alone, without their adjacency lists. The vector store is then free to place vectors by query locality rather than by vertex identity, which is what the next mechanism exploits.

Building on this decoupled architecture, we further reshape the in-memory graph hierarchy to reduce the memory cost of dynamic search. {\ourmethod} keeps the HNSW base layer on disk and retains only a sparse upper hierarchy, compact routing sketches, and caches in memory. To make this small in-memory state sufficient, {\ourmethod} groups the base layer into sections, where each section holds the vectors reached through the same upper-layer entry point. Over each section it builds an SA-tree sketch~\cite{navarro2002spatial} that keeps only the top of the tree. Routing on a sketch runs entirely in memory and selects an entry point close to the query in the on-disk base layer. Because the sketch carries part of the routing work, the in-memory HNSW hierarchy can remain small.

Finally, we introduce HNSW-based reordering to keep the separated vector path query-efficient. HNSW insertion already reveals which section a vector belongs to, because descent reaches an upper-layer entry point before the vector is linked into the base layer. Vectors in the same section are likely to be evaluated together during refinement, so {\ourmethod} places their vector data on nearby pages. This reduces scattered vector reads without a separate clustering phase or a global reordering pass.

The second limitation is addressed by the architecture as a whole rather than by any single mechanism. Both paths absorb updates out of place on disk, so memory no longer holds the dynamic state, and the compact sketches make a small in-memory hierarchy sufficient for routing. HNSW-based reordering then keeps vector reads page-efficient once the two paths are separated.

\vspace{1mm}
\noindent\textbf{Contributions.}
We make the following contributions in this paper.

\begin{itemize}
    \item We present {\ourmethod}, a disk-based dynamic vector search system built on a decoupled architecture. {\ourmethod} separates graph navigation from vector access, so graph connections and vector data can follow storage layouts matched to their different read and write behavior.
    \item To keep a dynamic HNSW base layer on disk, {\ourmethod} stores it in a graph-oriented LSM-tree that absorbs edge changes out of place. On top of this store, we design SA-tree-based routing, which uses compact in-memory sketches to select base-layer entry points close to the query and reduces graph-side read amplification.
    \item We introduce HNSW-based reordering, a vector-data layout strategy that extracts locality information from HNSW insertion and descent. It places vector data with shared HNSW-derived locality on nearby disk pages, without a separate clustering or global reordering pass.
    \item We evaluate {\ourmethod} against FreshDiskANN and SPFresh on streaming SIFT and SPACEV workloads with up to a hundred million vectors. {\ourmethod} serves the same streams with up to 6.2\(\times\) less total memory footprint and 7.6\(\times\) less disk than the two baselines, and writes up to 15\(\times\) fewer bytes per insertion than SPFresh. At the hundred-million scale, it sustains the full update stream on a single server while keeping its tail latency stable.
\end{itemize}

\section{Background}

\vspace{1mm}
\noindent\textbf{Vector search and recall.}
Let \(P=\{x_1,\ldots,x_n\}\) be a collection of \(d\)-dimensional vectors, and let \(q\) be a query vector. We use \(D(q,x)\) to denote a dissimilarity measure between \(q\) and \(x\), such as Euclidean distance or cosine distance. Exact \(k\)-nearest-neighbor search returns the \(k\) vectors in \(P\) with the smallest values of \(D(q,x)\). Approximate nearest-neighbor search reduces query cost by returning an approximate size-\(k\) result set \(R_k(q)\). Given the exact top-\(k\) result set \(T_k(q)\), recall is defined as
\(
    \operatorname{Recall@k}(q)
    =
    \frac{\left|R_k(q)\cap T_k(q)\right|}{k}.
\)

\vspace{1mm}
\noindent\textbf{Graph-based ANN search.}
A graph-based ANN index represents vectors as vertices in a proximity graph \(\mathcal{G}=(V,E)\). Each vertex is connected to a bounded set of nearby or navigationally useful vertices. Starting from one or more entry vertices, graph search explores the index by repeatedly selecting a promising candidate, reading its adjacency list, evaluating the distances of newly discovered neighbors, and updating a bounded search frontier. This process allows the query to examine only a small portion of the collection.

HNSW~\cite{malkov2018efficient} organizes the proximity graph into multiple layers. Sparse upper layers provide coarse navigation toward the region containing vectors close to the query, while the denser base layer performs a more extensive local search. Each vertex keeps at most \(M\) links per layer, where \(M\) is the degree bound. A vertex's maximum layer is drawn from an exponentially decaying distribution whose normalization is conventionally tied to \(M\), so layer \(\ell\) holds about \(n/M^{\ell}\) vertices and each layer is about \(M\) times smaller than the one below it. A query enters at the top layer, greedily reaches a nearest vertex at each layer, and continues from that vertex at the layer below, so each layer hands the next one an entry point. For a vertex \(v\), let \(z(v)\) denote its embedding and \(N_\ell(v)\) denote its adjacency list at layer \(\ell\). We write \(N(v)\) when the layer is clear from context. Search alternates between two access operations: reading \(N_\ell(v)\) to enumerate candidate neighbors and reading their vector data to evaluate their distances to the query. Graph search therefore reads adjacency lists and vector data through separate accesses.

\vspace{1mm}
\noindent\textbf{HNSW insertion.}
An HNSW insertion first draws a maximum layer for the new vertex \(x\). The algorithm descends through the hierarchy, searches for candidate neighbors at each participating layer, and selects a bounded set of connections for \(x\). It then adds reciprocal connections from selected neighbors to \(x\). If an existing adjacency list exceeds its degree bound, neighbor pruning removes one or more of its edges.

Consequently, an insertion modifies state at two different granularities. The new vertex contributes one embedding \(z(x)\) and one or more adjacency lists \(N_\ell(x)\). The same insertion may additionally modify the adjacency lists of several existing vertices through reciprocal linking and pruning, while their embeddings remain unchanged. These graph updates are typically small, irregular, and distributed across the neighborhood reached during insertion.

\vspace{1mm}
\noindent\textbf{LSM-trees absorb small updates out of place.}
A log-structured merge tree (LSM-tree)~\cite{oneil1996lsm} buffers writes in memory and flushes them to disk as immutable sorted files, so an update becomes an append rather than an in-place rewrite. Background compaction later merges these files, discarding superseded versions and keeping a lookup to a few files. The design therefore suits workloads dominated by many small, scattered writes, which is the pattern HNSW insertion produces on adjacency lists. Graph-oriented LSM-trees such as Aster~\cite{mo2025aster} apply it to adjacency data by storing individual edge updates and merging them into per-vertex adjacency entries during compaction.


\section{The Design of {\ourmethod}}

\subsection{System Overview}
\label{sec:overview}

\begin{figure}[t]
\centering
\includegraphics[width=\columnwidth]{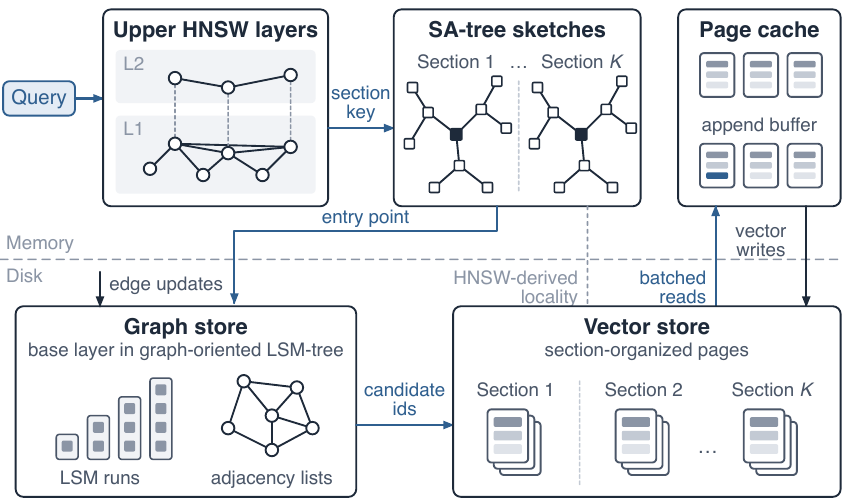}
\caption{Overview of {\ourmethod}.}
\vspace{-3mm}
\label{fig:lsmvec_overview}
\end{figure}

Section~\ref{sec:intro} leaves three requirements for a dynamic vector search system on disk: graph updates must not rewrite stable vector data, separation must not force queries onto a second search path or into a large resident state, and both on-disk paths must stay query-efficient. {\ourmethod} meets them with the layout in Figure~\ref{fig:lsmvec_overview}. The two access paths of the index live in separate stores, and memory keeps only routing state and caches. The HNSW base layer and the vector data both live on disk, the first in a graph-oriented LSM-tree and the second in a paged store. Memory holds a sparse upper HNSW hierarchy, compact routing sketches, and the caches and buffers that serve the two disk stores.

Separating the two stores raises a question the layout alone does not answer. Once vector placement is no longer tied to the graph layout, nothing keeps the vectors a query evaluates near one another. {\ourmethod} answers this with \emph{sections}. HNSW insertion already descends through the upper layers before a vector joins the base layer, and the entry point it reaches identifies a region of the base graph. We take that entry point as a locality key. The vectors reached through one entry point form a section, each section owns one routing sketch, and each section owns a set of vector pages. Routing and placement therefore follow the same key, and the insertion path supplies it at no extra cost.

Around this abstraction, the system has five components. In memory, it keeps the upper HNSW layers, the section-local SA-tree sketches, and the page cache with its append buffers. On disk, it keeps the graph store and the vector store.

\vspace{0.5mm}
\noindent\underline{\textit{Upper HNSW layers.}} These layers provide coarse navigation: they carry each query down to the entry point of the section where search continues. Because the sketches take over directional routing from there, these layers no longer carry the full routing work, so their layer ratio can be set aggressively and their resident size falls geometrically (Section~\ref{sec:memory}).

\vspace{0.5mm}
\noindent\underline{\textit{Section-local SA-tree sketches.}} Each sketch turns that entry point into a base-layer entry inside the query's neighborhood, before any disk access. Routing needs only a direction, and obtaining directions from the disk-resident base layer pays I/O at every inspected neighbor, whereas a tree descent follows parent-to-child choices in memory and never revisits a direction. Each sketch adapts a spatial approximation tree~\cite{navarro2002spatial}, whose nodes are data points rather than abstract cells, so a descent ends at a base-layer vertex from which refinement starts directly. A sketch keeps only the top hops of its tree, a bounded prefix much smaller than the section graph it routes over (Section~\ref{sec:memory}).

\begin{figure}[t]
\centering
\includegraphics[width=\columnwidth]{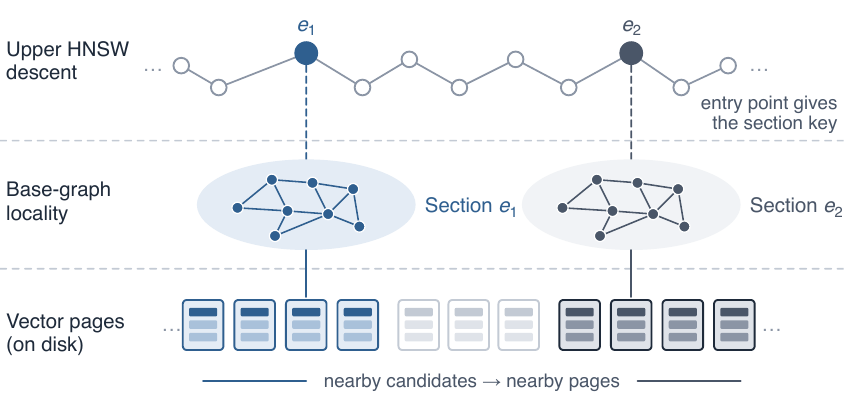}
\caption{HNSW-based reordering. Vectors assigned to the same Layer-1 entry point are stored on nearby vector pages.}
\label{fig:hnsw_reordering}
\vspace{-2mm}
\end{figure}

\vspace{0.5mm}
\noindent\underline{\textit{Graph store.}} This store holds the HNSW base layer in a graph-oriented LSM-tree, whose out-of-place, edge-granular writes match how HNSW changes its graph. An insertion adds a few edges across several adjacency lists, and the store absorbs each as an append, so stable vector data is never rewritten (Section~\ref{sec:structure}).

\vspace{0.5mm}
\noindent\underline{\textit{Vector store.}} This store holds the vector data on pages owned by sections, and resolves each vector id to its \((page,slot)\) location through a compact directory, so placement can follow section locality instead of id order. Reads inside one section then touch \(\lceil|S|/r\rceil\) pages rather than one page per candidate, so a single page read serves many distance evaluations (Section~\ref{sec:query}).

\vspace{0.5mm}
\noindent\underline{\textit{Page cache and append buffers.}} These serve reads and absorb inserts for the vector store. Their capacity is a configuration choice rather than a consequence of the design (Section~\ref{sec:memory}).

Taken together, memory serves routing and caching alone, so the resident budget is set by the design rather than by the volume of updates the index has absorbed. Section~\ref{sec:memory} bounds each part.

The rest of this section follows the components in the order a workload meets them. Section~\ref{sec:structure} describes the four index structures among them and how they are built, leaving the page cache and the append buffers, which serve those structures rather than hold the index, to Section~\ref{sec:memory}. Section~\ref{sec:query} follows a query through them, and Section~\ref{sec:update} follows an insertion, a deletion, and the background maintenance they trigger. Section~\ref{sec:memory} accounts for the memory the design keeps resident.

\subsection{Index Structure}
\label{sec:structure}

This subsection describes the four structures named above, beginning with the two that hold the index on disk, because the two resident structures exist to serve them. The hierarchy and the sections it defines are described together, since the section layer is part of the hierarchy.

\vspace{1mm}
\noindent\textbf{Matching HNSW's edge-granular updates to a graph-oriented LSM-tree.}
HNSW graph data are structurally irregular. Vertex degree is bounded, but reciprocal linking and diversity pruning leave adjacency lists of different lengths, so a uniform fixed-length record is a poor fit. The update pattern is equally irregular. Inserting a vertex writes its own adjacency list, and because HNSW links are reciprocal, the same insertion also writes one edge into the adjacency list of each selected neighbor, so one logical insertion becomes many single-edge writes scattered across the neighborhood it joined. Reads have the opposite shape, since refinement reads the whole adjacency list of each visited vertex. Many small scattered writes together with whole-object reads is the access pattern a graph-oriented LSM-tree is built for, so we store the HNSW base layer in Aster~\cite{mo2025aster}, which writes edge updates out of place and lets compaction merge them into vertex-based adjacency entries for queries. Aster is existing work. What matters here is that HNSW's update pattern matches what it does well, which is what lets the base layer live on disk without a read-modify-write path in the foreground.

Separating the two stores also reduces the data an insertion writes. A coupled record would have to be read and written back in full for each of the up to \(M\) existing adjacency lists an insertion modifies, and each such record carries a full \(d\)-dimensional vector that the insertion never changes. At \(M=16\), one insertion therefore moves up to sixteen stable vectors in order to record a change that affects only edges. The decoupled path appends the changed edges and nothing else.

\vspace{1mm}
\noindent\textbf{Packing fixed-length vector records into section-owned pages.}
Vector data have the opposite character. Every record is the same size, so the vector store can be a plain paged file in which each 4KB page holds a fixed number \(r\) of records. A page read thus brings in \(r\) vectors at once, and it pays off only when they are vectors a query evaluates together. We let sections rather than id ranges own pages, so each arriving vector is placed by its section key rather than its id, an online \emph{HNSW-based reordering} of the vector file (Figure~\ref{fig:hnsw_reordering}). The directory maps each id to its \((page,slot)\) location, so a reordered vector is still found in one lookup.

\vspace{1mm}
\noindent\textbf{Thinning the hierarchy until it carves the base layer into sections.}
The upper HNSW layers keep their usual role of coarse navigation, but not their usual size. In a standard hierarchy the layer-size ratio \(\rho\) is tied to the degree bound \(M\), because greedy graph routing needs enough vertices per layer to keep moving in the right direction. We instead set \(\rho\) far above \(M\) and keep a correspondingly sparse scaffold in memory. The section layer is Layer~1, the layer immediately above the base layer (Layer~0), regardless of how many navigational layers remain above it. Layer~1 holds one vertex per \(\rho\) base-layer vertices on average, so each of its vertices anchors a section of about \(\rho\) members, and when an insertion descends through it, the nearest vertex found there becomes the new vector's section key (Figure~\ref{fig:hnsw_reordering}). A vertex that later joins Layer~1 opens a new section for the insertions that follow, while earlier vectors keep the key their descent produced. The key therefore costs nothing beyond the descent, and it is available before the vector is written, which is what allows the graph store, the section sketch, and the vector pages to be updated under one key. The ratio thus sets the average section size, and with it how many pages a section spans and how much of it one sketch has to cover. Section~\ref{sec:eval-sens} sweeps this ratio, and Section~\ref{sec:memory} gives the memory saving. What permits a ratio far beyond \(M\) is the routing sketch that each section carries, introduced next.

\begin{algorithm}[t]
\caption{Construct a section's SA-tree.}
\label{alg:section_sa_tree}
\begin{algorithmic}[1]
\Require section members \(R\), root vertex \(r\), branching cap \(b\)
\State \(Q \leftarrow [(r,R)]\)
\While{\(Q\) is not empty}
    \State remove one pair \((v,R_v)\) from \(Q\)
    \State \(P \leftarrow R_v\setminus\{v\}\)
    \State \(L \leftarrow P\), sorted by increasing \(D(v,\cdot)\)
    \State \(C \leftarrow \emptyset\)
    \For{each \(x\in L\)}
        \State \textbf{if} \(|C|=b\), \textbf{break}
        \State \textbf{if} \(D(c,x)>D(v,x)\) for all \(c\in C\), add \(x\) to \(C\)
    \EndFor
    \State \textbf{if} \(C=\emptyset\), \textbf{continue}
    \State initialize \(R_c\leftarrow\{c\}\) for each \(c\in C\)
    \State \(U \leftarrow R_v\setminus(\{v\}\cup C)\)
    \State assign each \(x\in U\) to \(R_{\arg\min_{c\in C}D(c,x)}\)
    \State set the covering radius of each \(c\in C\) to \(\max_{x\in R_c}D(c,x)\)
    \State add edges \(v\rightarrow c\) for all \(c\in C\)
    \State append \((c,R_c)\) to \(Q\) for all \(c\in C\)
\EndWhile
\State \Return section SA-tree rooted at \(r\)
\end{algorithmic}
\end{algorithm}

\begin{figure}[t]
\centering
\includegraphics[width=0.70\columnwidth]{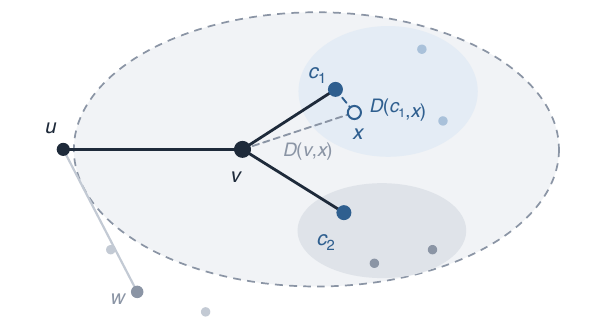}
\caption{Child admission in Algorithm~\ref{alg:section_sa_tree}. The dashed region holds the members assigned to \(v\), which broadly follow the direction of the parent edge from \(u\).}
\vspace{-3mm}
\label{fig:sa_admission}
\end{figure}

\vspace{1mm}
\noindent\textbf{Compressing each section into a tree-based routing sketch.}
We build one sketch per section by adapting the spatial approximation tree~\cite{navarro2002spatial} into a bounded-degree, depth-capped router rooted at the section's entry point. Algorithm~\ref{alg:section_sa_tree} gives the construction. At a node \(v\), candidates are considered in increasing distance from \(v\), and a candidate \(x\) becomes a child only if every child \(c\) already chosen satisfies \(D(c,x)>D(v,x)\). The remaining points are assigned to their nearest chosen child and queued as the next subproblem. The test admits children lying in different directions from \(v\) rather than the merely closest ones, which is what lets a descent pick a direction without scanning a dense neighborhood, and the branching cap \(b\) bounds how many directions one node exposes. Indeed, under Euclidean distance any two children sit more than sixty degrees apart around \(v\), because the admission test makes the segment between them the longest side of the triangle they form with \(v\), and the largest angle of a triangle faces its longest side. Each node stores a vector id, child pointers, and a covering radius.

For example, Figure~\ref{fig:sa_admission} follows one admission step. Because \(v\) and its siblings sit in well-separated directions around \(u\), and each of them takes the members nearest to it, the members left to \(v\) tend to extend the direction of the edge \(u\rightarrow v\). Among them, the nearest candidate becomes the first child \(c_1\), the second-nearest \(x\) lies closer to \(c_1\) than to \(v\) and is assigned to its region instead, and \(c_2\) is admitted because \(c_1\) does not cover its direction. A branch therefore tends to extend along the direction it inherits, and forks where the members spread beyond it.

{\ourmethod} does not keep the whole tree. Each additional level multiplies the prefix by up to \(b\) while the directional guidance is concentrated near the root, so we treat the full tree as a transient artifact of construction and keep only its root-to-\(H\)-hop prefix in memory, together with the vectors needed to compare a query against those nodes. Section~\ref{sec:query} routes over this prefix, and Section~\ref{sec:memory} shows what it costs.

Because these sketches carry the directional routing inside each section, the hierarchy above them can hold its ratio far beyond \(M\) without losing routing quality. This is what allows the large ratio introduced above, and Section~\ref{sec:memory} shows how much memory it saves.

\subsection{Query Processing}
\label{sec:query}

\begin{figure}[t]
\centering
\includegraphics[width=\columnwidth]{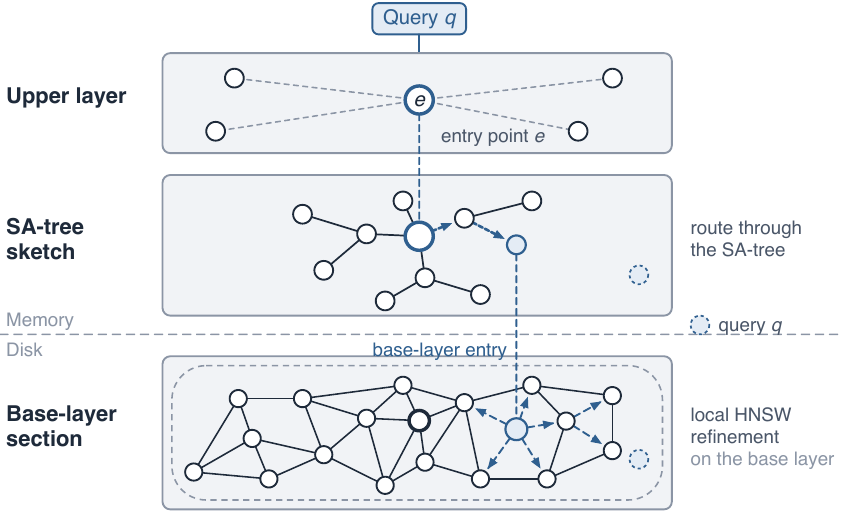}
\caption{SA-tree routing over one section. The descent runs in memory and selects a base-layer entry close to the query, from which refinement proceeds on disk.}
\vspace{-3mm}
\label{fig:sa_routing}
\end{figure}

This subsection follows a query through the structures just described. The query first descends the sparse upper layers to the entry point of a section, then routes over that section's tree-based sketch in memory, and enters the base layer on disk from the entry the sketch selects. From there, HNSW refinement produces candidate ids, and the vector store finally evaluates their distances with batched page reads. Figure~\ref{fig:sa_routing} illustrates the graph side of this path.

\vspace{1mm}
\noindent\textbf{Routing over the sketch before any I/O.}
Routing runs entirely on the in-memory prefix. Specifically, starting at the sketch root, the query compares itself against the children of the current node, follows the most promising directions, and descends at most \(H\) hops. Every comparison is a memory access, so routing adds CPU work but no disk reads, and all disk reads of a query happen during refinement. The descent ends at a base-layer vertex, which becomes the entry for the on-disk search.

\vspace{1mm}
\noindent\textbf{Refining on disk inside the answer's neighborhood.}
Refinement is standard HNSW search over the base layer, with the graph store serving adjacency lists. Unlike routing, every step here can reach disk: expanding a vertex reads its adjacency list from the graph store, and evaluating its candidates reads vector pages. Because the entry already sits inside the query's neighborhood, these reads are spent expanding vertices around the answer rather than crossing the graph toward it. Throughout, candidates are returned only after deleted ids are filtered by the live-node predicate, so a stale sketch entry can still guide routing without ever appearing in a result.

\vspace{1mm}
\noindent\textbf{Turning candidate evaluations into batched page reads.}
Each expansion discovers neighbors whose distances must be evaluated. To serve them, the vector store translates the candidate ids through the directory, groups them by page, and reads each touched page once, evaluating every candidate that resides on it. HNSW-based reordering is what makes this grouping productive. Specifically, a batch of \(m\) reads from one section touches at most \(\min(m,\lceil|S|/r\rceil)\) pages, a count that the section's own extent bounds however large the batch, and one read can serve up to \(r\) evaluations. For example, sixty candidate evaluations that fall inside one section are read from that section's own page range rather than from unrelated pages spread across the file. Moreover, refinement concentrates its expansions in a few sections once routing has succeeded, so the same pages are touched repeatedly within a query, and the page cache extends this reuse across nearby queries.

\vspace{1mm}
\noindent\textbf{Keeping every query on a single search path.}
The path above is the only one a query takes. Recent insertions are met in place, since their edges live in the graph store and their vector data in the section pages or the append buffer, and deletions are filtered at result time. Queries therefore never consult a second index, and search behavior does not change as updates accumulate.

\subsection{Update and Maintenance}
\label{sec:update}

\begin{figure}[t]
\centering
\includegraphics[width=\columnwidth]{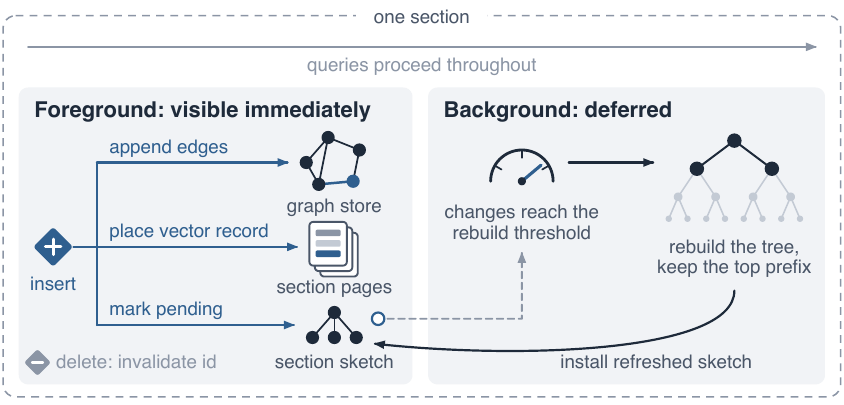}
\caption{Update handling within one section. Foreground operations become visible immediately, and accumulated changes trigger a deferred section-local rebuild.}
\label{fig:dynamic_maintenance}
\end{figure}

Whereas a query only reads these structures, an update extends them. A new vector brings a node and its edges to the graph, an embedding record to its section's pages, and a member to its section's sketch. {\ourmethod} makes each addition visible immediately and defers everything else. A foreground operation records what queries must see and returns, while section-local maintenance keeps each section's sketch fresh in the background. Figure~\ref{fig:dynamic_maintenance} shows this split for one section.

\vspace{1mm}
\noindent\textbf{Foreground updates touch one graph neighborhood and one section.}
An insertion follows the descent described in Section~\ref{sec:structure}. Specifically, the descent yields the section key, the new edges are appended to the graph store, and the vector data enters a reclaimed slot of the same section when one exists or the section's append buffer otherwise, with the directory updated to the new location. The affected sketch records the new member as pending. Nothing in this path reorders other sections or rebuilds any structure, so the foreground cost of an insertion is bounded by one graph neighborhood and one section.

\vspace{1mm}
\noindent\textbf{Making deletion a visibility change.}
A deletion must take effect in results immediately. From the moment it is issued, the deleted vector may not appear in any answer, so {\ourmethod} invalidates the id and the live-node predicate excludes it from every subsequent query. Everything else about a deletion is a scheduling choice, and {\ourmethod} schedules it in the background, as other dynamic indexes do. We deliberately retain a deleted vector's sketch entry until the section's next rebuild, because its geometry still guides routing toward its neighborhood, and it counts toward the rebuild threshold, so a section that accumulates deletions is rebuilt sooner. The vector's slot is likewise reclaimed by a later insertion into the same section. Queries thus observe every deletion immediately, while the foreground path stays as short as an insertion's.

\vspace{1mm}
\noindent\textbf{Amortizing sketch refresh with section-local rebuilds.}
Over time, pending members and deleted entries accumulate in each section. Let a section hold \(n_S\) live members with \(p_S\) accumulated changes. {\ourmethod} rebuilds the section's sketch when \(p_S \geq \alpha n_S\), where \(\alpha\) is the rebuild threshold. A rebuild materializes the full section tree transiently, extracts the fresh top-\(H\) prefix, refreshes the routing vectors, and discards the rest. Charging the rebuild to the changes that triggered it gives an amortized cost of at most \(C_{\mathrm{build}}(n_S)/(\alpha n_S)\) per change, so a larger section pays more per change only through its build cost. For example, at \(\alpha=0.3\) a section absorbs changes amounting to nearly a third of its live members before its next rebuild, and that rebuild touches this section alone. A smaller \(\alpha\) refreshes routing more eagerly, whereas a larger one trades staleness for fewer rebuilds.

\vspace{1mm}
\noindent\textbf{Confining concurrency to section boundaries.}
Every write is an append or a slot write inside one section, so none rewrites a structure another operation may be reading. An insertion installs the vector record and its directory entry before appending any edge that can reach the new vertex. A deletion publishes through id invalidation alone, so its records stay in place, traversal skips the invalidated vertex, and the live-node predicate keeps it out of every result. Maintenance observes the same boundary. A rebuild consumes the pending and stale state of one section and installs the refreshed sketch when it finishes, leaving every other section untouched, which is what lets it run behind foreground work.

\subsection{Memory Budget}
\label{sec:memory}

Memory is where the preceding design choices converge. Section~\ref{sec:intro} observed that existing dynamic vector search systems on disk hold in-memory state that grows with the index and its update volume, and set the standard a better design should meet: memory reserved for compact routing state and caches, whose size is a configuration choice. {\ourmethod} meets this standard by keeping the index data itself on disk, so neither the base layer nor the vector store is held in memory. What remains is the upper hierarchy, one sketch per section, and the working state of the vector store, and this subsection bounds each in turn.

\vspace{1mm}
\noindent\textbf{Shrinking the upper hierarchy geometrically.}
Under the geometric layer model, Layer~\(\ell\) holds \(|V_\ell| \approx n/\rho^{\ell}\) vertices for an index of \(n\) vectors, so an upper hierarchy whose vertices keep \(O(M)\) links holds
\[
\sum_{\ell\ge1}|V_\ell|\,M \approx Mn\sum_{\ell\ge1}\rho^{-\ell} = O\!\left(\frac{Mn}{\rho-1}\right)
\]
state in memory. Increasing \(\rho\) therefore shrinks the hierarchy at a geometric rate. Section~\ref{sec:structure} established what licenses a ratio far above \(M\): the tree-based sketches carry the directional routing that the removed layers would otherwise perform.

\vspace{1mm}
\noindent\textbf{Bounding each sketch by a constant-size prefix.}
A sketch with branching cap \(b\) and depth \(H\) holds at most
\[
C_H(b)=\frac{b^{H+1}-1}{b-1}
\]
nodes, whatever the size of its section. With \(n/\bar{s}\) sections of average size \(\bar{s}\), the sketches together hold at most \((n/\bar{s})\,C_H(b)\) nodes, against \(O(Mn)\) edge endpoints for the section graphs they route over, a ratio of \(O(C_H(b)/(M\bar{s}))\). Since \(\bar{s}\approx\rho\) by construction, thinning the hierarchy also thins this ratio. A sketch also never holds more nodes than its section has members, so it stays within \(1/M\) of the edge endpoints it routes over. In the benchmark setting, \(b=4\) and \(H=4\) give \(C_H(4)=341\), and at \(M=16\) each sketch holds at most one sixteenth as many nodes as its section graph holds edge endpoints.

\vspace{1mm}
\noindent\textbf{Capping the working state of the vector store.}
The remaining in-memory state serves the vector store. The page cache and the section append buffers have fixed capacities, and capacity governs I/O rather than result quality. Specifically, the disk stores hold the authoritative data, so a smaller cache costs additional page reads, and a faster update stream drains the buffers more often instead of enlarging them. Alongside them, the id-to-location directory resolves each vector id to the page that holds its record, at a small fixed cost per vector.

Taken together, the upper hierarchy shrinks geometrically in \(\rho\), each sketch is bounded by \(C_H(b)\) nodes whatever its section holds, the directory costs a small fixed amount per vector, and the caches and buffers stay at fixed capacity. None of these terms scales with the volume of updates absorbed, which is the quantity Section~\ref{sec:intro} identified as the one existing designs cannot bound. Section~\ref{sec:eval-e2e} measures the resulting footprint.

\section{Related Work}

\vspace{1mm}
\noindent\textbf{Traditional and graph-based ANN indexes.}
Approximate nearest-neighbor search has been widely studied through tree-based, hash-based, quantization-based, and graph-based indexes. Classical trees and randomized partitions provide compact routing structures~\cite{silpa2008optimised,bernhardsson2017annoy,li2023learning}, while locality-sensitive hashing trades memory and probes for probabilistic recall~\cite{charikar2002similarity,huang2015query}. Quantization and inverted-file designs reduce distance-evaluation cost through compressed candidates~\cite{jegou2010product,guo2020accelerating,johnson2019billion,facebook2020faiss}, and layout designs accelerate the distance evaluation itself~\cite{Kuffo_2025}. Modern high-performance ANN search is largely graph-based. NSW and HNSW introduced navigable small-world graphs~\cite{malkov2014approximate,malkov2018efficient}, and constructions such as NSG and NGT refine edge selection and construction quality~\cite{aoyama2011fast,fu2017fast,iwasaki2015ngt,iwasaki2018optimization,manohar2024parlayann}. Surveys and benchmarks document this space extensively~\cite{li2019approximate,wang2021comprehensive,Azizi_2025,aumuller2020ann}. Another line studies routing and convergence properties of proximity graphs, including pruning rules and monotonicity conditions~\cite{harwood2016fanng,fu2022ssg,peng2023efficient,indyk2023worstcase,lu2024probabilistic,Wang_2024,li2025fastconvergent}. Most of these approaches assume that the index is built and searched in memory, which becomes expensive as workload scale increases.

\vspace{1mm}
\noindent\textbf{Disk-based vector indexes.}
The growth of modern unstructured data workloads has pushed vector indexes beyond DRAM~\cite{zhang2018zoom,ren2020hm,gong2020idec,chen2021spann}. DiskANN is the canonical disk-based approach. It builds a pruned Vamana graph offline, keeps compressed vectors in memory to order candidates, stores the graph together with full-precision vectors on disk, and batches neighborhood reads with beam search~\cite{jayaram2019diskann}. SPANN clusters vectors into posting lists to reduce random reads~\cite{chen2021spann}. Recent systems further improve on-disk search performance by optimizing disk layout, page granularity, and SSD-aware query execution, including Starling~\cite{wang2024starling}, PageANN~\cite{anon2027pageann}, PipeANN~\cite{guo2026pipeann}, and MARGO~\cite{yue2025margo}. However, these systems assume an index built offline over a fixed collection and provide no update path. A streaming workload therefore cannot run on them, and the evaluation compares against dynamic disk-based systems instead.

\vspace{1mm}
\noindent\textbf{Dynamic ANN maintenance.}
Real-world vector search increasingly requires ANN indexes to ingest new data without rebuilding the full structure. One line studies dynamic ANN in memory. GTI combines graph and tree structures to support logarithmic updates~\cite{Ma_2024}. Other systems maintain monotonic paths, support distributed streaming search, or serve update-heavy vector databases~\cite{Liu_2025,Gong_2025,Hu_2025}. MIRAGE-ANNS improves graph construction while preserving insertion support~\cite{voruganti2025mirage}. These works are relevant to dynamic ANN, but they either assume primarily in-memory index state or target a different construction problem, so they sit outside the on-disk setting this paper studies.
The closest on-disk dynamic systems are FreshDiskANN and SPFresh. FreshDiskANN keeps recent updates in short-term in-memory indexes and periodically integrates them into a long-term index on disk, so a query searches both structures until the merge completes~\cite{singh2021freshdiskann}. SPFresh instead applies updates to a primary partitioned index on disk, where each partition groups nearby vectors and is represented by an in-memory summary. Its LIRE protocol reorganizes affected partitions incrementally, so that they remain efficient to search as new vectors arrive~\cite{xu2023spfresh}. They are the baselines in our evaluation. Greator speeds up FreshDiskANN's merge by rewriting only the affected disk blocks with multiple update threads~\cite{yu2025greator}. OdinANN inserts vectors directly into the on-disk graph and keeps search stable under concurrent inserts, relying on in-memory quantized vectors for navigation and pruning~\cite{guo2026odinann}. Their gains come from quantization and multi-threaded execution, which are orthogonal to our design, and both keep each vector in the same on-disk record as its adjacency list.

\vspace{1mm}
\noindent\textbf{Vector search systems.}
Beyond index algorithms, vector database systems study how vector search is served inside data systems, including storage management and filtered retrieval~\cite{guo2022manu,Patel_2024}. RAG serving systems place vector retrieval inside end-to-end generation pipelines~\cite{Jiang_2024,Jiang_2025}. Another line accelerates vector retrieval with heterogeneous execution on GPUs and FPGAs~\cite{zhang2024fast,tian2025towards,zeng2023df}. These works are complementary to our focus on dynamic on-disk ANN index maintenance, and {\ourmethod} can serve as the index layer beneath them.

\section{Evaluation}

\begin{figure*}[!t]
\centering
\includegraphics[width=\textwidth]{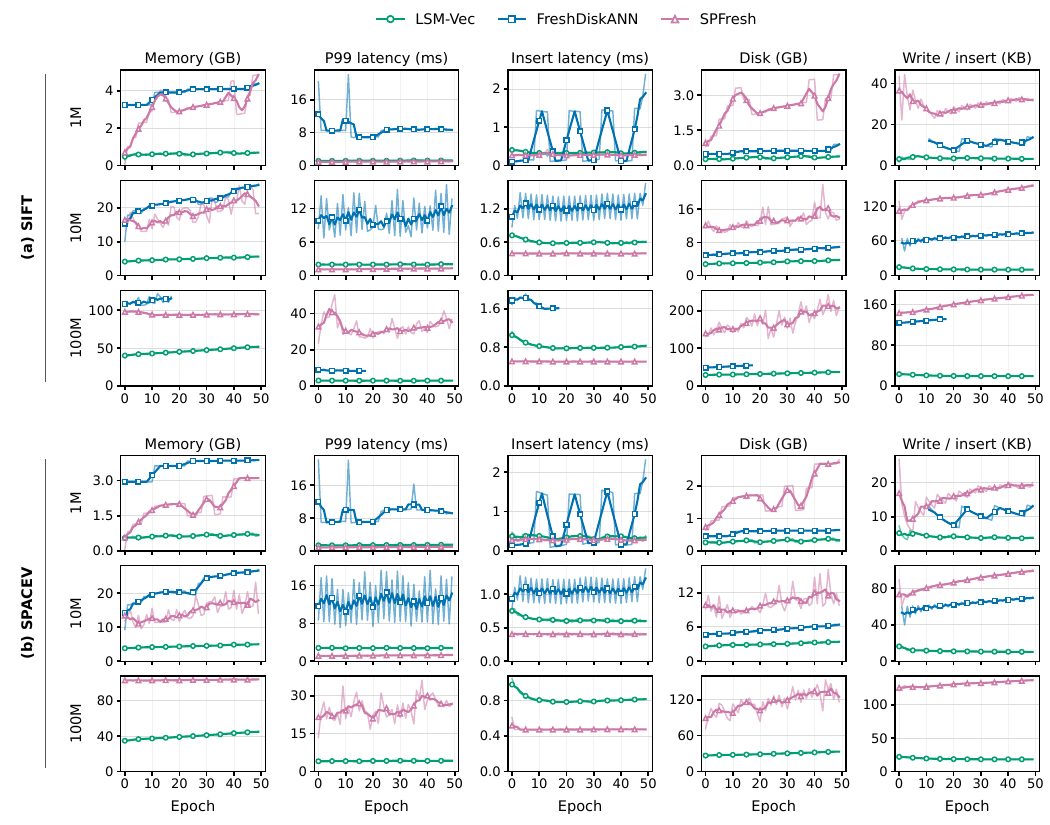}
\caption{End-to-end behavior under the R-9010 streaming workload as the collections grow. Rows are SIFT and SPACEV at 1M, 10M, and 100M. Columns report total memory footprint, P99 query latency, insert latency, disk footprint, and bytes written per insertion over epochs. Light lines show per-epoch values, and bold lines show their 5-epoch moving averages.}
\label{fig:eval-summary}
\end{figure*}

This section evaluates {\ourmethod} on streaming workloads for on-disk ANN search. The main experiments use SIFT and SPACEV from 1M to 100M vectors and compare {\ourmethod} with FreshDiskANN and SPFresh. Section~\ref{sec:eval-e2e} reports the end-to-end comparison as the collections grow. Section~\ref{sec:eval-ablation} then isolates the two read-side mechanisms, and Section~\ref{sec:eval-sens} varies the graph degree and the layer-size ratio.

\subsection{Experimental Setup}

\vspace{1mm}
\noindent\textbf{Datasets.}
We evaluate on the two vector collections that both dynamic baselines are evaluated on.

\vspace{0.5mm}
\noindent\underline{\textit{SIFT/BIGANN}}~\cite{jegou2011searching} contains 128-dimensional unsigned-byte image descriptors under Euclidean distance, with a query set of 10,000 vectors.

\vspace{0.5mm}
\noindent\underline{\textit{SPACEV}}~\cite{contributors2021spacev1b} is derived from data of a commercial search engine and represents a different form of encoding, 100-dimensional signed-byte deep natural-language vectors under Euclidean distance. We sample its query set deterministically to 10,000 vectors to match SIFT's query cardinality.

\vspace{1mm}
Both collections are widely used for evaluating dynamic vector search, and they approach it from hybrid angles. Evaluating each at 1M, 10M, and 100M gives six workload cells that span both data types and two orders of magnitude in scale.

\vspace{1mm}
\noindent\textbf{Workloads.}
Each experiment starts from a base index of \(N\) vectors and then applies a streaming update trace for 50 epochs. The R-9010 workload inserts \(0.9\%N\) vectors and deletes \(0.1\%N\) vectors per epoch, using a disjoint insertion pool of \(0.5N\) vectors. All systems run on the same byte-identical update trace within each dataset and scale. We recompute exact top-100 ground truth over the active set every 10 epochs and report recall@10 at those checkpoints.

\vspace{1mm}
\noindent\textbf{Systems.}
We compare {\ourmethod} with two disk-based dynamic baselines. FreshDiskANN represents an out-of-place disk graph that periodically merges recent updates into the on-disk index. We use its merge-enabled configuration, which triggers StreamingMerge every 100K updates. SPFresh represents an in-place dynamic partitioned index with local rebalancing. Both systems read through the file system, which is how vector databases are commonly deployed in practice. We therefore run SPFresh on its file-I/O backend rather than the user-space stack it also supports. All three systems keep full vectors in each collection's native byte type, so they index the same data at the same precision. Both baselines run at the settings their authors report, with a search candidate list of \(L=150\) for FreshDiskANN and a search budget of \(ef=64\) for SPFresh. {\ourmethod} keeps the HNSW base layer and vector data on disk and routes queries with the in-memory top-\(H\) sketches described in Section~\ref{sec:structure}. Unless stated otherwise, we use \(M=16\), layer-size ratio \(\rho=256\), branching cap \(b=4\), sketch depth \(H=4\), and rebuild threshold \(\alpha=0.3\).

\vspace{1mm}
\noindent\textbf{Platform.}
All experiments run on an Ubuntu 22.04.5 Linux server with an Intel Core i9-13900K CPU, 125GB RAM, and a 2TB Western Digital PC SN810 NVMe SSD formatted with ext4. The machine exposes 32 hardware threads. All benchmarks are implemented in C++ and compiled with g++ 10.5.0.

\vspace{1mm}
\noindent\textbf{Measurements.}
All systems use four threads for base-index construction and one foreground thread for the streaming workload. Background flush and compaction threads are left as configured in each system's released code, including the RocksDB settings in SPFresh's code. Queries are issued serially, so reported latency is per-query latency rather than aggregate multi-client throughput. We report two memory metrics. \emph{Own memory} is the anonymous resident memory of the serving process, measured as RssAnon. It covers everything the process allocates, including index state, caches, and buffers, and it excludes the workload data and replay bookkeeping that each benchmark driver holds. \emph{Total memory footprint} adds the OS page cache that the system's reads keep resident, measured by periodically scanning the store files with mincore and, where such scans miss cached pages, by system memory statistics. The distinction matters because a buffered-read system can serve from cache pages it does not own. The operating-system caching each system depends on is therefore counted the same way for all three. The end-to-end comparison runs {\ourmethod} at the insertion search budget \(efc=12\) (HNSW's efConstruction) with the query budget \(ef_{\mathrm{final}}=64\).

\subsection{End-to-End Results}
\label{sec:eval-e2e}

\begin{figure}[t]
\centering
\includegraphics[width=\linewidth]{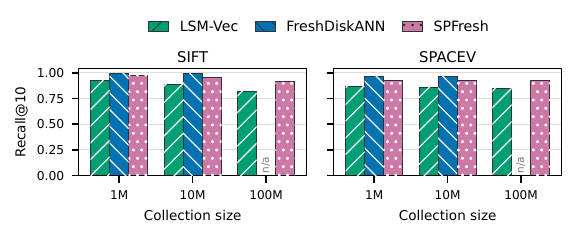}
\caption{Recall@10 across scales, averaged over the R-9010 stream.}
\label{fig:recall-scales}
\end{figure}

\begin{figure*}[t]
\centering
\includegraphics[width=0.85\textwidth]{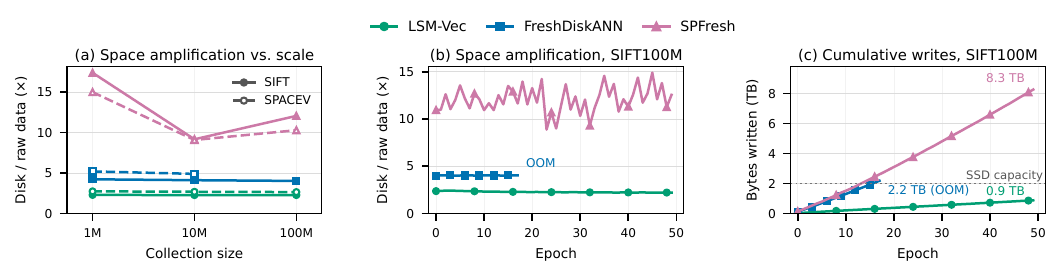}
\caption{Storage cost under the R-9010 stream. (a) Disk footprint divided by the raw size of the live vectors, averaged over the stream, at 1M, 10M, and 100M. (b) The same ratio per epoch on SIFT100M. (c) Cumulative bytes written on SIFT100M. FreshDiskANN stops after 18 epochs at this scale, when it runs out of memory.}
\label{fig:storage-cost}
\end{figure*}

Figure~\ref{fig:eval-summary} reports the six workload cells across five measures: total memory footprint, P99 query latency, insert latency, disk footprint, and bytes written per insertion. Light lines plot the per-epoch values and bold lines their 5-epoch moving averages, so the gap between the two shows how steady a system is from one epoch to the next. {\ourmethod}'s per-epoch values stay close to its trend in every measure, whereas SPFresh's memory, disk footprint, and P99 latency swing noticeably between epochs as its background storage maintenance proceeds. Figure~\ref{fig:recall-scales} reports recall averaged over the stream, and Figure~\ref{fig:storage-cost} summarizes the storage cost.

\vspace{1mm}
\noindent\textbf{The memory gap holds as the collections grow.}
Figure~\ref{fig:eval-summary} (first column) reports the total memory footprint, which adds the page cache a system's reads keep resident to the memory it allocates itself. Averaged over the stream, {\ourmethod} needs 0.6GB at 1M, about 5GB at 10M, and 40--46GB at 100M, 3.1--4.9\(\times\) less than SPFresh and 4.5--6.2\(\times\) less than FreshDiskANN at 1M and 10M. The curves also differ in how they grow. {\ourmethod}'s footprint grows in step with the live set, which expands by 40\% over the stream, because its resident state is fixed by configuration rather than by the updates it absorbs. SPFresh's footprint grows faster than its data, rising three- to fivefold over the 1M streams, and FreshDiskANN's climbs as its in-memory delta accumulates, reaching 104GB of own memory at SIFT100M before the machine runs out. At 100M, SPFresh's footprint stops growing only because its store, 115--210GB on disk, has outgrown what memory can hold. {\ourmethod} still stays 2.1\(\times\) below it on SIFT100M and 2.6\(\times\) below it on SPACEV100M. The next paragraph shows how this affects SPFresh's query latency.

\begin{table}[t]
\centering
\small
\setlength{\tabcolsep}{3.2pt}
\begin{tabular}{lcccc}
\toprule
 & \multicolumn{3}{c}{Latency, mean / P99 (ms)} & \makecell{Device reads} \\
\cmidrule(lr){2-4}
System & SIFT1M & SPACEV1M & SIFT10M & \makecell{per query, 10M} \\
\midrule
{\ourmethod} & \textbf{0.87} / \textbf{1.34} & 0.99 / \textbf{1.57} & \textbf{2.71} / \textbf{5.70} & \textbf{90KB} \\
SPFresh & 1.11 / 3.07 & \textbf{0.97} / 2.25 & 3.69 / 9.55 & 489KB \\
FreshDiskANN & 7.67 / 8.84 & 6.14 / 7.31 & 5.71 / 6.79 & -- \\
\bottomrule
\end{tabular}
\caption{Query latency with all caches dropped before each query phase, including the OS page cache of the store, and warmed with 2,000 pool vectors.}
\label{tab:cold-cache}
\end{table}

\vspace{1mm}
\noindent\textbf{Comparison of query performance.}
Figure~\ref{fig:eval-summary} (second column) shows that {\ourmethod}'s smaller memory does not come at the cost of query stability. We report P99 latency, because a dynamic index is judged by how its query path behaves while updates accumulate, and instability appears in the tail first. {\ourmethod}'s tail stays nearly flat over every stream and grows only mildly with scale, from 1.2--1.4ms at 1M to 2.0--2.8ms at 10M and 2.9--4.1ms at 100M, 3--8\(\times\) below FreshDiskANN's wherever both run. This growth is far slower than that of the collection. From 1M to 100M, the vector pages a query reads from the store grow from 303 to 673 on SIFT and from 381 to 781 on SPACEV, about 2.1\(\times\) for a hundredfold larger collection, because each query still descends to a few sections and searches locally there. SPFresh behaves differently, as the memory analysis above suggests. At 1M and 10M its store still fits in memory, and both systems answer within a few milliseconds. At 100M its store grows to 115--210GB, more than memory can hold, so the postings a query scans must increasingly be read from disk. Its P99 rises more than twentyfold from 10M, to 25--33ms on average with peaks of 50ms, and its mean latency keeps rising through the stream as the store grows. At 100M, {\ourmethod}'s P99 is 6--11\(\times\) lower than SPFresh's. Two mechanisms sustain this behavior. SA-tree routing enters the base layer close to the target region, and batched page reads amortize the remaining disk accesses. Both mechanisms operate on the read path itself, so {\ourmethod} reaches this latency without holding large index state in memory.

\begin{figure}[t]
\centering
\includegraphics[width=\linewidth]{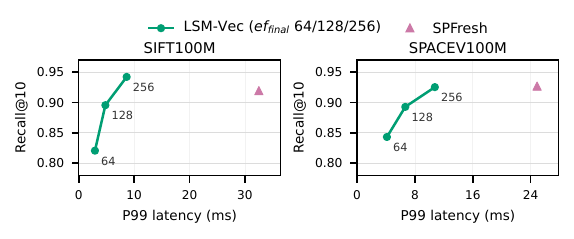}
\caption{Recall versus P99 query latency on SIFT100M and SPACEV100M, averaged over the R-9010 stream. {\ourmethod} is shown at final search budgets of 64, 128, and 256, and SPFresh at its default setting. FreshDiskANN is omitted because it does not complete the stream at this scale.}
\label{fig:budget-100m}
\end{figure}

\vspace{1mm}
\noindent\textbf{{\ourmethod} depends less on cached pages.}
The query phases above replay the same query set in every epoch, so each phase runs on caches warmed by the one before it. Table~\ref{tab:cold-cache} repeats the SIFT1M, SPACEV1M, and SIFT10M runs with every cache dropped before each query phase, including the OS page cache of the store, and warmed only with 2,000 vectors from the insertion pool. Recall is unchanged. {\ourmethod} then has a lower P99 than both baselines in every run, 1.34ms against SPFresh's 3.07ms on SIFT1M and 5.70ms against 9.55ms on SIFT10M. SPFresh depends more on cached pages: at 10M it reads 489KB per query from the disk, against 90KB for {\ourmethod}.

\vspace{1mm}
\noindent\textbf{{\ourmethod} remains efficient at hundred-million scale.}
At 100M the R-9010 stream grows the active set to 140 million vectors over the 50 epochs. {\ourmethod} completes the stream on both datasets with its tail latency and insert cost as stable as at the smaller scales. Across the 50 epochs, its P99 stays between 2.7 and 3.0ms on SIFT100M and between 3.8 and 4.3ms on SPACEV100M, while SPFresh's ranges from 24 to 50ms on SIFT100M, and each insertion writes 16--23KB, against 116--210KB for SPFresh. In contrast, FreshDiskANN's in-memory state exhausts the machine's memory after 18 epochs on SIFT100M and cannot complete its build on SPACEV100M. Figure~\ref{fig:budget-100m} examines the query side by varying {\ourmethod}'s final search budget. A larger budget raises recall at a modest cost in latency. At \(ef_{\mathrm{final}}=256\), {\ourmethod} reaches 0.942 on SIFT100M and 0.925 on SPACEV100M, matching or exceeding SPFresh, at a P99 of 8.6ms and 10.8ms, 2.3--3.8\(\times\) below SPFresh's. At every budget we measured, {\ourmethod} answers faster than SPFresh, so the application can choose the budget according to its recall target.

\begin{figure}[!t]
\centering
\includegraphics[width=\linewidth]{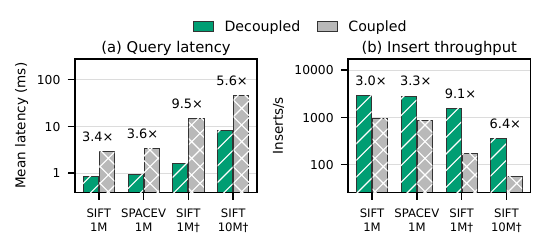}
\caption{Decoupled stores versus a coupled variant that keeps each base-layer vertex as one LSM record, under the R-9010 stream. \(\dagger\) marks runs with the process memory capped after the build, at 650MB for SIFT1M and 3.5GB for SIFT10M. Labels give the advantage of the decoupled design.}
\label{fig:coupled}
\vspace{-3mm}
\end{figure}

\begin{figure}[t]
\centering
\includegraphics[width=\linewidth]{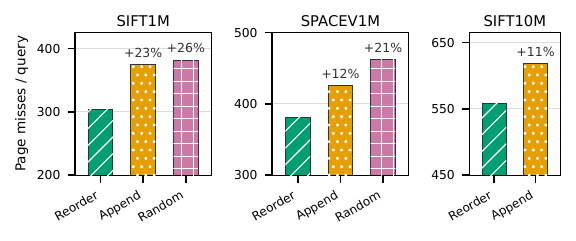}
\caption{HNSW-based reordering ablation on SIFT1M, SPACEV1M, and SIFT10M: vector-page misses per query. The variants change only the vector placement policy.}
\label{fig:reordering-ablation}
\vspace{-3mm}
\end{figure}

\vspace{1mm}
\noindent\textbf{{\ourmethod} achieves near-best insert latency.}
Figure~\ref{fig:eval-summary} (third column) reports insert latency. {\ourmethod} inserts in 0.3--0.8ms on average, and its latency stays flat across each stream. An insertion appends its adjacency changes to the graph store and writes one record into its section pages, two independent appends. FreshDiskANN's foreground insertions touch only its in-memory delta, but the delta must eventually be merged into the disk graph, and the merge blocks the insertions waiting behind it, causing spikes of up to 7ms. SPFresh's foreground latency is in the same sub-millisecond range, because much of its write work is done by background storage threads. These threads run during the insertions and take CPU time: measured as process CPU time over wall time, SPFresh uses 1.21 cores while inserting on SIFT10M and 1.37 on SIFT100M, against 1.03 for {\ourmethod} on SIFT10M. That work appears in the bytes written per insertion (fifth column). An in-place posting update rewrites whole postings, including vector bytes the insertion does not change, so SPFresh writes 5--15\(\times\) more bytes per insertion than {\ourmethod}, and FreshDiskANN writes 3--7\(\times\) more. Over the SIFT100M stream, SPFresh writes 8.3TB in total, against 0.9TB for {\ourmethod} (Figure~\ref{fig:storage-cost}(c)). Section~\ref{sec:structure} predicted this asymmetry from the layout alone.

\vspace{1mm}
\noindent\textbf{{\ourmethod} stays compact on disk.}
Figure~\ref{fig:storage-cost} divides each system's disk footprint by the raw size of the vectors it indexes. {\ourmethod} stays at 2.3--2.7\(\times\) at every scale and has the smallest store in every cell of Figure~\ref{fig:eval-summary} (fourth column), because compaction reclaims the space that updates leave behind. FreshDiskANN stays at 4--5\(\times\). SPFresh needs 9--17\(\times\), because its released configuration assigns each vector to up to eight postings near partition boundaries and stores every copy in full, and at SIFT100M its store grows to 210GB, against 37GB for {\ourmethod}. {\ourmethod} therefore stays small in memory and compact on disk at the same time.

\begin{figure}[t]
\centering
\includegraphics[width=\linewidth]{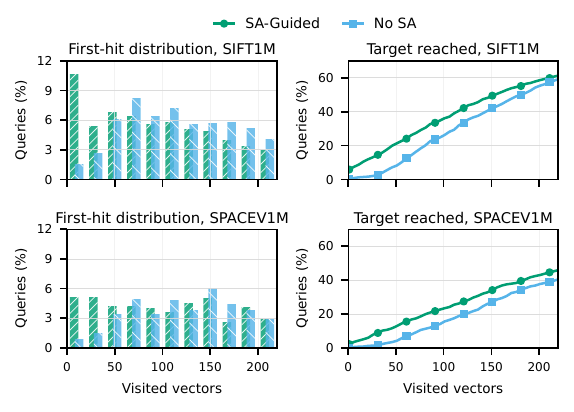}
\caption{SA-tree routing case study on SIFT1M and SPACEV1M, at a fixed base-layer search budget.}
\label{fig:routing-case-study}
\vspace{-3mm}
\end{figure}

\subsection{Ablations}
\label{sec:eval-ablation}

We first isolate the decoupled architecture itself and then the two read-side mechanisms, one per access path, so the end-to-end efficiency can be attributed to identifiable design components rather than to the operating point.

\vspace{1mm}
\noindent\textbf{Decoupling speeds up both queries and updates.}
Figure~\ref{fig:coupled} compares {\ourmethod} with a coupled variant that stores each base-layer vertex as one LSM record holding both its adjacency list and its vector, with everything else unchanged. Without a memory limit, coupling slows queries by 3.4--3.6\(\times\) and inserts by 3.0--3.3\(\times\) at equal recall, and it writes 3.4--4.4\(\times\) more bytes per insertion. More than half of the bytes a coupled query fetches belong to adjacency lists it never expands, and a graph-only update must rewrite the vector stored beside each list it modifies. The gap widens once memory is capped. At 650MB on SIFT1M, the coupled variant is 9.5\(\times\) slower to query and 9.1\(\times\) slower to insert, and at SIFT10M under a 3.5GB cap the gaps remain 5.6\(\times\) and 6.4\(\times\). Separating the stores lets each path read and write only the data it needs, which matters most in the memory-constrained regime that {\ourmethod} targets. The graph store adds little read amplification of its own. Even when its LSM-tree grows to three levels at SIFT10M, a graph lookup reads about one on-disk file on average, and this cost stays flat across the stream as compaction reshapes the tree.

\vspace{1mm}
\noindent\textbf{HNSW-based reordering reduces page reads.}
Figure~\ref{fig:reordering-ablation} isolates HNSW-based reordering by holding the graph and the query budget fixed and changing only where vector records are written, by section key under \emph{Reorder}, in arrival order under \emph{Append}, or at unrelated pages under \emph{Random}. The sweep is available only because the two stores are separate, since a layout that binds a vector to its adjacency list cannot place the two by different rules. All three layouts touch a similar number of vector pages per query, but they differ in how many of those pages must be read from the store. Arrival order and random placement incur 23\% and 26\% more page misses than Reorder on SIFT1M, and 12\% and 21\% more on SPACEV1M, at the same recall. At 10M the gap persists: on SIFT10M, where we compare Reorder with Append only, Append incurs 11\% more page misses at the same recall. Once graph search concentrates around a local region, placing vectors from the same HNSW-derived section on nearby pages lets one page read serve several of them. The saving comes from the layout alone. Reordering adds no query-time computation and no separate clustering pass, so it applies to every query at no additional cost.

\vspace{1mm}
\noindent\textbf{Sketch routing enters the base layer closer to the target.}
Figure~\ref{fig:routing-case-study} gives the graph-side counterpart under one base-layer budget, where \emph{SA-Guided} enters the base layer through the section's sketch and \emph{No SA} descends into it directly. The first-hit distribution shows that under SA-Guided, one query in nine reaches its exact nearest neighbor within the first 20 visited vectors, against fewer than one in fifty without the sketch, whose hits peak only after 60 visits. The difference remains later in the search: 21\% of queries reach the target within 50 visits, against 8\% without the sketch, and the median path to the target is 16\% shorter. SPACEV1M shows the same pattern: 13\% of queries reach the target within 50 visits, against 4\% without the sketch, and the median path is 12\% shorter. A closer entry lets refinement reach the target with fewer expansions, which is where the graph-side page savings originate, so the sketch improves routing quality in addition to bounding memory.

\begin{figure}[!t]
\centering
\includegraphics[width=\linewidth]{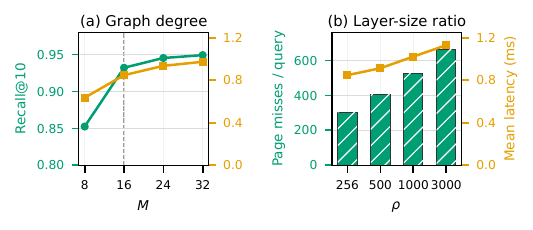}
\caption{Sensitivity to the graph degree and the layer-size ratio on SIFT1M, averaged over the R-9010 stream. (a) reports recall and mean query latency, with the default \(M=16\) marked. (b) reports vector-page misses per query and mean query latency.}
\label{fig:sensitivity}
\vspace{-3mm}
\end{figure}

\subsection{Parameter Sensitivity}
\label{sec:eval-sens}

\vspace{1mm}
\noindent\textbf{Sensitivity to graph degree and layer ratio.}
Figure~\ref{fig:sensitivity} varies the two parameters that shape the graph and the hierarchy. The graph-degree sweep places the recall knee at \(M=16\). Halving the degree to 8 costs eight points of recall, while doubling it to 32 adds under two points and raises mean query latency by 15\%. The layer-size ratio acts mainly on reads. As \(\rho\) falls from 3000 to 256, sections become smaller and more numerous, so the vectors a query evaluates concentrate on fewer pages. Page misses per query drop by 54\% and mean latency by 25\%, at a small cost in total memory, while stream-mean recall stays within one point across the sweep. We therefore set \(\rho=256\). That is still sixteen times the degree bound, so the resident hierarchy remains far sparser than in a standard HNSW.

\section{Conclusion}

This paper presents {\ourmethod}, a dynamic disk-based ANN system that keeps the upper HNSW layers in memory, stores the base layer on disk, and absorbs updates out of place. Two mechanisms keep it query-efficient, SA-tree-based routing for compact entry selection and HNSW-based reordering for section-aware layout. On streaming SIFT and SPACEV workloads, {\ourmethod} substantially reduces the memory footprint relative to FreshDiskANN and SPFresh while remaining competitive in recall, latency, and update throughput.

\bibliographystyle{ACM-Reference-Format}
\bibliography{IEEEabrv,sample}

\end{document}